\documentclass{aa}  %

\usepackage{graphicx}
\usepackage{txfonts}
\usepackage{xcolor}
\begin{document}

   \title{Planetary Nebulae of the Large Magellanic Cloud II: the connection with
    the progenitors' properties}

%  \subtitle{I. Gas }

   \author{P. Ventura\inst{1,2}, S. Tosi\inst{3,4}, D. A. Garc{\'\i}a-Hern{\'a}ndez\inst{5,6},
           F. Dell'Agli\inst{1}, D. Kamath\inst{7}, \\ L. Stanghellini\inst{8}, 
           S. Bianchi\inst{3}, M. Tailo \inst{9}, M. A. Gómez-Muñoz\inst{10,11}}

   \institute{INAF, Observatory of Rome, Via Frascati 33, 00077 Monte Porzio Catone (RM), Italy \and 
              Istituto Nazionale di Fisica Nucleare, section of Perugia, Via A. Pascoli snc, 06123 Perugia, Italy \and
              Dipartimento di Matematica e Fisica, Università degli Studi Roma Tre, 
              Via della Vasca Navale 84, 00100, Roma, Italy \and 
              LNF, Laboratori Nazionali Fascati, Via Enrico Fermi, 54, 00044 Frascati Roma, Italy \and
              Instituto de Astrof\'isica de Canarias (IAC), E-38205 La Laguna, Tenerife, Spain \and
              Departamento de Astrof\'isica, Universidad de La Laguna (ULL), E-38206 La Laguna, Tenerife, Spain \and
              School of Mathematical and Physical Sciences, Macquarie University, Balaclava Road, Sydney, NSW 2109, Australia \and
              NSF's NOIRLab, 950 Cherry Ave., Tucson, AZ 85719, USA \and
              Osservatorio Astronomico di Padova, Vicolo dell'Osservatorio 5, 35122 Padova, Italy \and
              Departament de Física Quàntica i Astrofísica (FQA), Universitat de Barcelona (UB),  c. Martí i Franqués, 1, 08028 Barcelona, Spain \and
              Institut de Ciències del Cosmos (ICCUB), Universitat de Barcelona (UB), c. Martí i Franqués, 1, 08028 Barcelona, Spain.
              }

   \date{Received September 15, 1996; accepted March 16, 1997}

% \abstract{}{}{}{}{} 
% 5 {} token are mandatory

 \abstract
   {The study of planetary nebulae (PNe) offers the opportunity of evaluating the efficiency
    of the dust production mechanism during the very late asymptotic giant branch 
    (AGB) phases, which proves crucial to assess the role played by AGB stars as 
    dust manufacturers.}
   {We study the relationship between the properties of PNe, 
    particularly the gas and dust content, with the 
    mass and metallicity of the progenitor stars, to understand
    how dust production works in the late AGB phases, and to shed new
    light on the physical processes occurring to the stars and the material
    in their surroundings since the departure from the AGB until the PN
    phase.}
   {We consider a sample of 9 PNe in the Large Magellanic Cloud, 7 out 
     of which characterized by the presence of carbonaceous dust, the remaining
     2 with silicates. For these stars the masses and the metallicity of the 
     progenitor stars were estimated.
    We combine results from stellar evolution and dust formation
    modelling with those coming from the analysis of the spectral 
    energy distribution, to find the relation between the
    dust and gas mass of the PNe considered and the
    mass and metallicity of the progenitors.
    }
   {    
   The physical properties of carbon-rich PNe are influenced by the mass of the progenitor star. Specifically, the dust-to-gas ratio in the nebula increases from $5\times 10^{-4}$ to $6\times 10^{-3}$ as the progenitor star's mass increases from approximately $\rm 0.9~M_{\odot}$ to $\rm 2~M_{\odot}$. This change is partly influenced by the effective temperature of the PNe, and it occurs because higher-mass carbon stars are more efficient at producing dust. Consequently, as the progenitor's mass increases, the gas mass of the PNe decreases, since the larger amounts of dust lead to greater effects from radiation pressure, which pushes the gas outwards. No meaningful conclusions can be drawn by the study of the
    PNe with silicate-type dust, because the sub-sample is made up of 2 PNe only, one of which
    is almost dust-free.}
   {}


   \titlerunning{Characterizing Planetary Nebulae of the Large Macellanic Clouds}
   \authorrunning{Ventura et al.}
   \maketitle
%
%-------------------------------------------------------------------

\section{Introduction}
Over the past few decades, significant advancements have been made in understanding 
the evolution of stars passing through the AGB phase. In particular, these improvements 
pertain to changes in key physical properties and alterations in the surface chemical 
composition, which are closely linked to the progenitor stars' mass and metallicity 
\citep{karakas14,ventura22}.

Important further progresses in this field have been accomplished in the last decade, when some
research teams coupled self-consistently the results from stellar evolution
modelling with the chemo-dynamical description of the circumstellar envelope
\citep{ventura12, ventura14, nanni13, nanni14}, to derive the dust production 
rate (DPR) of the stars during the AGB phase. These studies allowed the
characterization of the evolved stellar populations of the Magellanic
Clouds \citep{flavia15a, flavia15b}, a few Local Group 
galaxies \citep{flavia16, flavia18b, flavia19} and Andromeda (Gavetti et al. 2024).

The target of these investigations extends beyond the comprehension of the still
poorly known aspects of the AGB evolution, the ambitious goal being to assess the
role that these objects play in the pollution of the interstellar medium,
with the gas expelled via stellar winds and the dust formed in significant
quantities in their expanded circumstellar envelope, where the physical
conditions prove extremely favourable to the condensation of gaseous 
molecules into dust grains \citep{gs99}. Progresses in this field are required
to understand the relative contribution from AGB stars to form the dust
currently observed in galaxies \citep{raffa14} and, more generally, to
establish the contribution from these stars in the dust budget of the
Universe \citep{raffa24}. 

The modelling of dust production in the wind of evolved stars
showed that in most cases the bulk of the dust is released during the late AGB phases. 
This is definitively the case for carbon stars, as the DPR increases steadily 
during the AGB evolution, owing to the accumulation of carbon in the surface regions, 
favoured by the effects of periodic third dredge-up (TDU) events \citep{ventura14}. 
Low-mass stars that fail to reach the C-star stage also follow this behaviour, as 
only during the very late AGB phases the temperatures in the external regions of the 
star and in the circumstellar envelope are cool enough to allow dust formation 
\citep{ventura22}. Recent studies \citep{ester23} suggested that this is the case also for 
$\rm M \geq 4~M_{\odot}$ stars experiencing hot bottom burning \citep[HBB,][]{sackmann92}.

The fact that most of the dust is released towards the end of the AGB evolution
stimulated the study of the phases following the AGB, in the attempt
of using the observations of post-AGB stars and PNe to reconstruct the chain of
events occurring since the very late AGB phases, when dust formation stopped: the 
target of these investigations is the determination of the efficiency of the
dust production mechanism at the tip of the AGB, when most of the dust is 
released, thus the most relevant for the determination of the dust yields. 

The initial progress in this field was made by \citet{tosi22}, who studied a 
sample of post-AGB stars in the Magellanic Clouds, revealing key insights into dust 
formation during the late AGB phase. This was followed by \citet{tosi23} and 
\citet{flavia23a}, who focused on Galactic post-AGB stars, further refining our 
understanding of their mass-loss histories and dust properties. These studies 
highlighted the potential to trace the physical conditions of the final AGB 
stages—particularly the intensity of dust production—by analyzing the infrared 
excess of post-AGB stars.

\citet{flavia23b} advanced the study of dust production by tracing the evolutionary 
path from the late AGB phase through the post-AGB phase to the PN stage. They proposed 
that this approach allows for a comprehensive assessment of the gas and dust mineralogy 
and quantity surrounding the central object. By analyzing the IR excess of a single star 
with a progenitor mass slightly above solar across different evolutionary stages, 
\citet{flavia23b} aimed to understand how dust disperses from the star after dust formation 
ceases and to estimate the DPR during these epochs. This study provides insights into dust 
and gas release during the late AGB phases and the subsequent PN stage.

Stimulated by the potentialities of the exploration of the AGB - post-AGB - PN
sequence to reconstruct the processes of gas release and dust production occurring
during the late AGB phases, we decided to expand the results obtained by \citet{flavia23b},
limited to a single object, to explore a wider sample of PNe, for which 
observations are available in a wide spectral region, extending from the UV to the
IR. A first paper (Tosi et al. 2024, hereafter referred to as Paper I) was dedicated
to the analysis of a sample of Large Magellanic Cloud (LMC) planetary nebulae: from 
detailed SED fitting analysis we derived the main chemical and dusty properties of 
the PNe and, by comparing with the evolutionary tracks across the HR diagram,
we inferred the mass and chemical composition of the progenitor stars. In the 
present work we follow the same approach proposed by \citet{flavia23b} to deduce
information on the dust production taking place during the late AGB phases and
the dynamical properties of the wind during the transition from the AGB to the
PN phase: to this aim, we interpret the information obtained in paper I on the
basis of results from stellar evolution and dust formation modelling regarding the
AGB phase of stars with the same masses derived in paper I for the individual
objects.

Overall, we intend to create a synergy
between the studies of dust production in the wind of AGBs, which are able
to predict the dust formed as a function of mass and metallicity, and the
analysis of PNe observations. The direct comparison of evolutionary models and 
observed properties would constrain the SED changes from the AGB to the initial 
white dwarf stellar phases, allowing us to classify the observed targets 
following the \citet{pedro} scheme, based on ISO spectra.

This paper is structured as follows: the ingredients used to model the stellar evolution
and the dust formation process, and to analyse the SED of the PNe, are described
in section \ref{input}; the properties of the stars believed to be progenitors of the
sources in the sample considered in the present work, are discussed in section \ref{evol};
section \ref{disc} is devoted to the analysis of the connection between the properties of 
stars and the nebulae of the PNe and their previous evolutionary history; finally,
the conclusions are given in section \ref{concl}.

\begin{table*}
\caption{Main properties of the PNe investigated in paper I, and of the progenitor stars: 
1. Source ID; 2-3. Luminosity and effective temperature of the CS; 
4. Mass of the gas in the nebula; 5. Dust-to-gas ratio of carbon or 
silicate dust (logarithmic units); 6-7. Mass and metallicity of the progenitor 
stars}
\label{tab1}      
\centering
%\addtolength{\leftskip}{+2cm}
\addtolength{\leftskip}{-0.5cm}
\begin{tabular}{c c c c c c c}    
\hline 
ID  &  L$/\rm{L}_\odot$   &  $\rm{T}_{\rm{eff}}$[K]  & $\rm{M}_{\rm{gas}}$$/~\rm{M}_\odot$  &  $\log(\delta$[C])  &  $\rm M_{init}/M_{\odot}$  & $Z$              \\ 
\hline
  Carbon dust
\\
\hline
\\
SMP LMC 4  &  6\,500 $^{+600 }  _{-100  } $  &  105\,000 $^{+5\,000 }_{-5\,000 }$  &  0.034 
$^{+0.002} _{-0.001}$ &  -2.20  $^{+0.07}_{-0.12}$   &  $1.5$  &  $8\times 10^{-3}$ \\ \\
SMP LMC 18   &  2\,000  $^{+700 }  _{-500  } $   &   50\,000  $^{+6\,000 }_{-5\,000 }$   &       0.086 $^{+0.020} _{-0.016}$ &   -3.61 $^{+0.38} _{-0.01}$  &  0.9  &  $2\times 10^{-3}$   \\ \\
SMP LMC 25   &  4\,900  $^{+400 }  _{-200  } $   &   60\,000  $^{+10\,000}_{-10\,000}$    &       0.110 $^{+0.003} _{-0.003}$ &   -2.58 $^{+0.15} _{-1.17}$  &  0.9  &  $2\times 10^{-3}$   \\ \\
SMP LMC 34   &  4\,500  $^{+800 }  _{-600  } $   &   46\,000  $^{+4\,000 }_{-3\,000}$  &       0.282 
$^{+0.009} _{-0.035}$ &   -3.58 $^{+0.21} _{-0.10}$  &  0.8  &  $2\times 10^{-3}$   \\ \\
SMP LMC 66   &  4\,500  $^{+2\,000} _{-1\,500 } $  &   107\,000 $^{+5\,000 }_{-7\,000}$   &       0.205 $^{+0.041} _{-0.028}$ &   -3.13 $^{+0.12} _{-0.13}$  &  1    &  $4\times 10^{-3}$   \\ \\
SMP LMC 71   &  5\,400  $^{+400 }  _{-100  } $   &   164\,000 $^{+9\,000 }_{-4\,000 }$   &        0.065 $^{+0.002} _{-0.005}$ &   -2.21 $^{+0.15} _{-0.10}$  &  2  &  $8\times 10^{-3}$     \\ \\
SMP LMC 102  &  3\,600  $^{+900 }  _{-800  } $   &   140\,000 $^{+9\,000 }_{-10\,000}$   &        0.370 $^{+0.042} _{-0.064}$ &   -3.52 $^{+0.12} _{-0.20}$  &  1.25  &  $4\times 10^{-3}$  \\ \\

\hline
  Silicates &             &            &         &    $\log(\delta$[Sil])                           &   &        \\
\hline
\\
SMP LMC 81   &  4\,700  $^{+1\,200}  _{-800  } $   &   80\,000  $^{+40\,000}_{-15\,000}$   &       $0.129^{+0.006} _{-0.009}$ & $-2.47^{+0.09}_{-0.10}$  &  0.8  & $4\times 10^{-3}$   \\ \\
\hline   
Dust free
\\
\hline
\\
SMP LMC 80   &   3\,200  $^{+900 }  _{-1\,100 } $   &   57\,000 $ ^{+5\,000 }_{-3\,000 }$  &      $0.051^{+0.007} _{-0.005}$ & $-$  &  0.6  & $4\times 10^{-3}$                       \\ \\
\hline
\label{tabpost}
\end{tabular}
\end{table*}

\section{Numerical modelling and physical assumptions}
\label{input}
In this study we focus on the PNe sources investigated in paper I, with the 
aim of linking the properties of the gas and dust in the nebulae, here summarized in
Table 1, with the evolutionary history of the progenitor stars, and the efficiency 
of the dust formation mechanism that characterised the late AGB phases. 
The interpretation of the results from paper I proposed in the present
work is based on the combination of findings from the modelling of the
AGB evolution of the stars believed to be the progenitors of the sources
considered, and of the dust formation process taking place during the AGB
phases of the same stars, particularly those close to the beginning of the 
post-AGB evolution. In this section we give a brief summary of the numerical 
and physical ingredients
adopted.

\subsection{Stellar evolution modelling}
\label{stevol}
The starting point of the interpretation of the results obtained in paper I is 
the computation of the evolutionary sequences of the model stars considered. We note that,
unlike paper I, here we are not only interested in the excursion of the evolutionary
tracks across the HR diagram, but also in the AGB variation of the physical quantities
that are essential ingredients to determine the efficiency of dust production, such
as the mass loss rate and the surface chemistry. 

To build the
evolutionary sequences we used the ATON code for stellar evolution, in the updated version
whose numerical structure is described in detail in \citet{ventura98}. The code is able 
to follow the entire evolution of low and intermediate mass stars, from the pre-MS, to 
the white dwarf cooling sequence.  Among others, here
we mention only the input physical ingredients of ATON most relevant for the present work,
which are: a) the mass-loss rate during the C-star phases are modelled according
to the description proposed by the Berlin group \citep{wachter02, wachter08}, whereas
for the oxygen-rich phases we adopted the treatment by \citet{blocker95}; 
b) the surface molecular opacities for mixtures enriched in the CNO elements are
calculated by means of the AESOPUS tool, described in \citet{aesopus}; c) the 
overshoot from convective borders is described by means of a diffusive approach,
where convective velocities are assumed to decay exponentially with an e-folding
distance of $\rm 0.002$ pressure scale heights (this is in agreement with the
calibration of the luminosity function of the LMC carbon stars given in
\citet{ventura14}).

The description of the model stars considered in the following sections is based on
extant evolutionary sequences previously published by our group: for what attains
the metallicities $Z=0.004$ and $Z=0.008$ we use results from \citet{marini21}, whereas 
the low-metallicity counterparts of $Z=0.002$ were published in \citet{devika23}. 

\subsection{Dust production by AGB stars}
\label{dustmod}
Dust formation in the wind is described following the approach
proposed by the Heidelberg team \citep{fg01, fg02, fg06}, which was used in
previous works by our group \citep{ventura12, ventura14}. For some
selected evolutionary stages (around 20) taken during each of the inter-pulses
experienced by the model stars calculated with the ATON code, according
to the description given in section \ref{stevol}, we model dust formation on the basis of the 
mass, effective temperature, luminosity, mass loss rate and surface
chemistry of the star. 

On the chemical side, the most relevant factor for dust 
formation is the surface $\rm C/O$, with oxygen-rich stars producing silicates and
alumina dust, whereas in carbon rich environments the formation of solid carbon
and silicon carbide takes place. This is due to the very large stability of the
CO molecule, which absorbs the least abundant element between carbon and oxygen in its entirety.
Therefore, in the case of carbon stars, the amount of solid
carbon dust that can form is given by the carbon excess with respect to oxygen.
The solution of the set of equations listed and discussed in \citet{ventura12}, which
govern the dynamic and thermodynamic stratification of the stellar winds and the growth of 
dust grains, allows us to evaluate the mineralogy of the dust formed and the
dust production rates of the various dust species considered.

\subsection{The interpretation of the observations of the planetary nebulae}
For completeness, we also briefly report here the road followed to interpret the 
observations given in paper I, where we used observed spectra 
and photometric data to model the SED of the PNe through the spectral synthesis code 
CLOUDY \citep[v22.02; ][]{ferland17}. To this aim, we built synthetic SED models, based on a 
detailed description of the CS's, the surrounding gaseous nebula, and the dust emission. 
To model the CS we used atmospheric models from \cite{rauch03} and \cite{pauldrach01}, scaling 
the chemical abundances from \cite{aller83} and \cite{khromov89} to align with the observed 
values from \citet{leisy06} and \citet{henry89}. 
The gaseous nebula was modeled under the assumption of spherical geometry, with a 
constant hydrogen density and an inner radius consistent with photometric measurements 
from \citet{shaw01}, as described in Paper I. 
This comprehensive modelling approach enabled the derivation of key PN properties 
such as the nebular gas mass $\rm{M}_{\rm{gas}}$ and the dust-to-gas mass ratios $\delta$. 
Further details on the SED modeling and results are reported in paper I.

\section{The evolution and dust production of low-mass AGB stars}
\label{evol}
The classification of the sources discussed in paper I was mainly based on the comparison
between their position on the HR diagram and the evolutionary tracks of stars
of different mass and metallicity, obtained with the ATON code, described in section
\ref{input}. The derived surface chemical composition, in particular the carbon mass fraction, 
was also used for a better identification of the mass and chemical composition of the progenitor 
stars (paper I). In this section we discuss the AGB evolution and dust production mechanism of 
the progenitors, to be able to relate the properties of the PNe, derived from the analysis of
the SED, with the physical processes that occurred during the final part of the AGB phase.

7 out of the 9 sources studied in paper I descend from carbon stars, with the exceptions of
SMP LMC 80 and SMP LMC 81, which are  oxygen-rich. The AGB evolution of the stars that reach
the C-star stage was discussed in a number of interesting reviews \citep{busso99, karakas14,
ventura22}. The upper limit of the initial mass of these stars is around $\rm 3~M_{\odot}$
\citep{ventura13}, as in higher mass objects the ignition of HBB at the base of the convective envelope
prevents the stars from becoming carbon stars. The afore-mentioned limit is partly sensitive 
to the metallicity, as HBB is ignited more easily in lower-metallicity environments \citep{flavia18a}.
The mass range of carbon stars is also
limited from below \citep{karakas14}, as reaching the C-star stage demands a minimum number of 
TDU events, which can be experienced only if the initial mass of the
envelope is above a minimum threshold. This threshold limit is also sensitive to the
chemical composition \citep{ventura22}, and decreases as the metallicity of the stars decreases: 
indeed, the evolution of low-metallicity stars favor the carbon star outcome, given their  
lower oxygen content\footnote{The lower limit of the mass of the stars that become carbon stars
is also sensitive to the amount of mass lost during the ascending of the
red giant branch: a large mass loss experienced during the red-giant branch (RGB) inhibits the formation
of carbon stars. This issue is relevant for $\rm M<1.5~M_{\odot}$ stars, as higher mass
objects lose only a small fraction of their envelope mass during the RGB phase. To
prevent any ambiguity, we will refer to the mass of the star at the start of the
core helium burning phase across the text.}.

As discussed in \citet{ventura22}, the evolution of carbon stars of a given metallicity is
primarily driven by the initial mass. The higher the initial mass, the higher the number of 
thermal pulses and TDU events experienced, the larger the amount of carbon accumulated in the
surface regions. The progeny of stars whose mass is near the HBB ignition threshold, i.e.
$\rm \sim 2.5-3~M_{\odot}$, reach surface carbon mass fractions of the order of $1\%$: under these
conditions the surface regions expand, owing to the notable increase in the low-T molecular
opacities \citep{marigo02}, which cause a general cooling of the external layers and the
expansion of the whole stellar structure. These conditions prove extremely favourable for the
condensation of gaseous molecules into solid particles, a process that increases the
rate of mass loss, thus shortening the evolutionary time scales. \citet{flavia15a}
proposed that the evolved stars in the LMC exhibiting the largest IR excesses,
populating the reddest side of the colour-colour plane obtained with the Spitzer
magnitudes, descend from progenitors in this mass range.

Generally speaking, the dust production rate of carbon stars increases since the start of the C-star phase, 
owing to the rise in the surface carbon. Therefore, most of the dust is produced and released during the 
latest inter-pulse phases, which are therefore the most relevant to assess the gas and dust pollution from 
these stars. \citet{marini21} suggested that all the reddest stars in the LMC, including part of the
objects labelled as EROS, are carbon stars evolving through the latest AGB phases.

\begin{figure*}
\vskip-40pt
\begin{minipage}{0.46\textwidth}
\resizebox{1.\hsize}{!}{\includegraphics{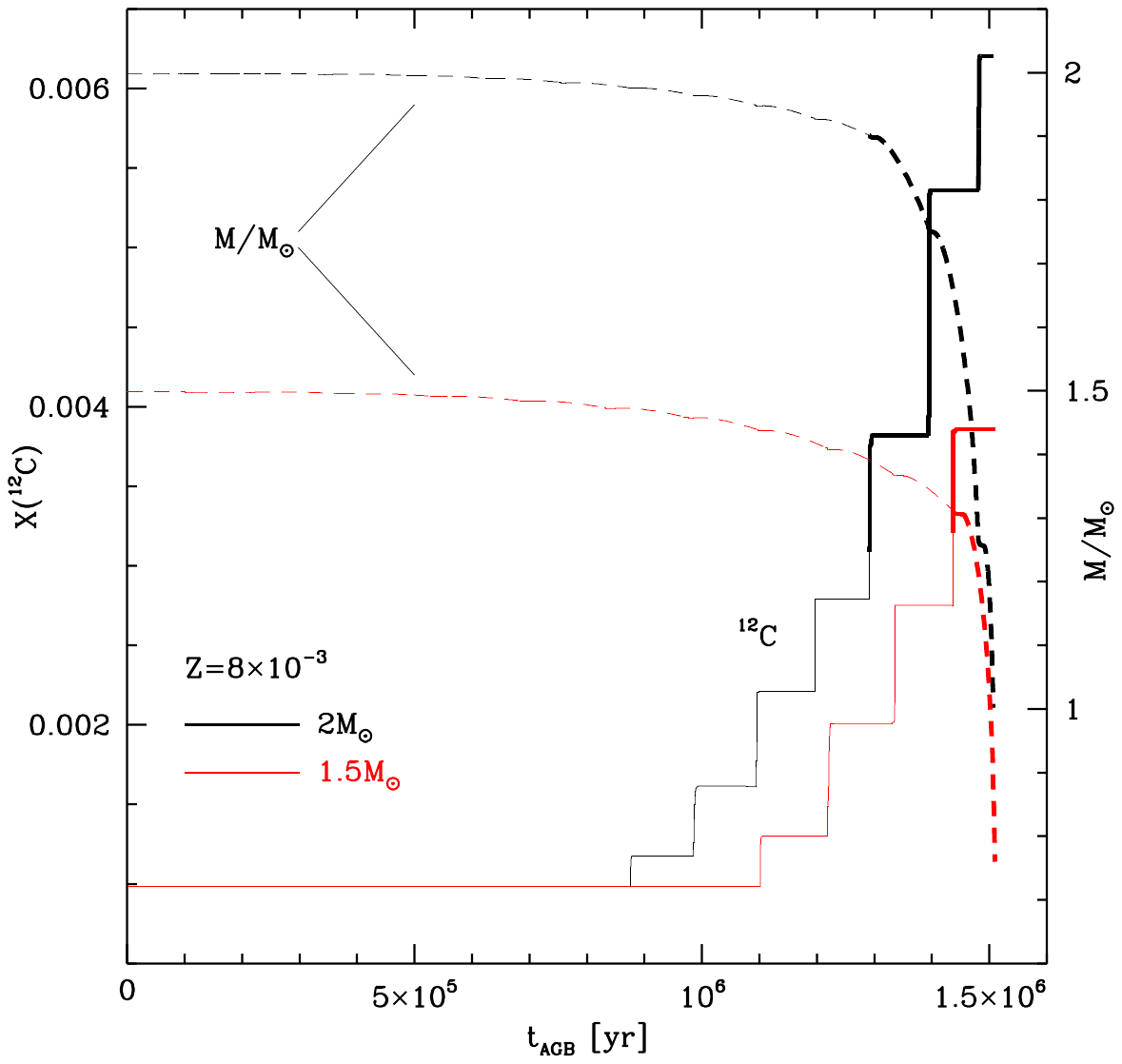}}
\end{minipage}
\begin{minipage}{0.46\textwidth}
\resizebox{1.\hsize}{!}{\includegraphics{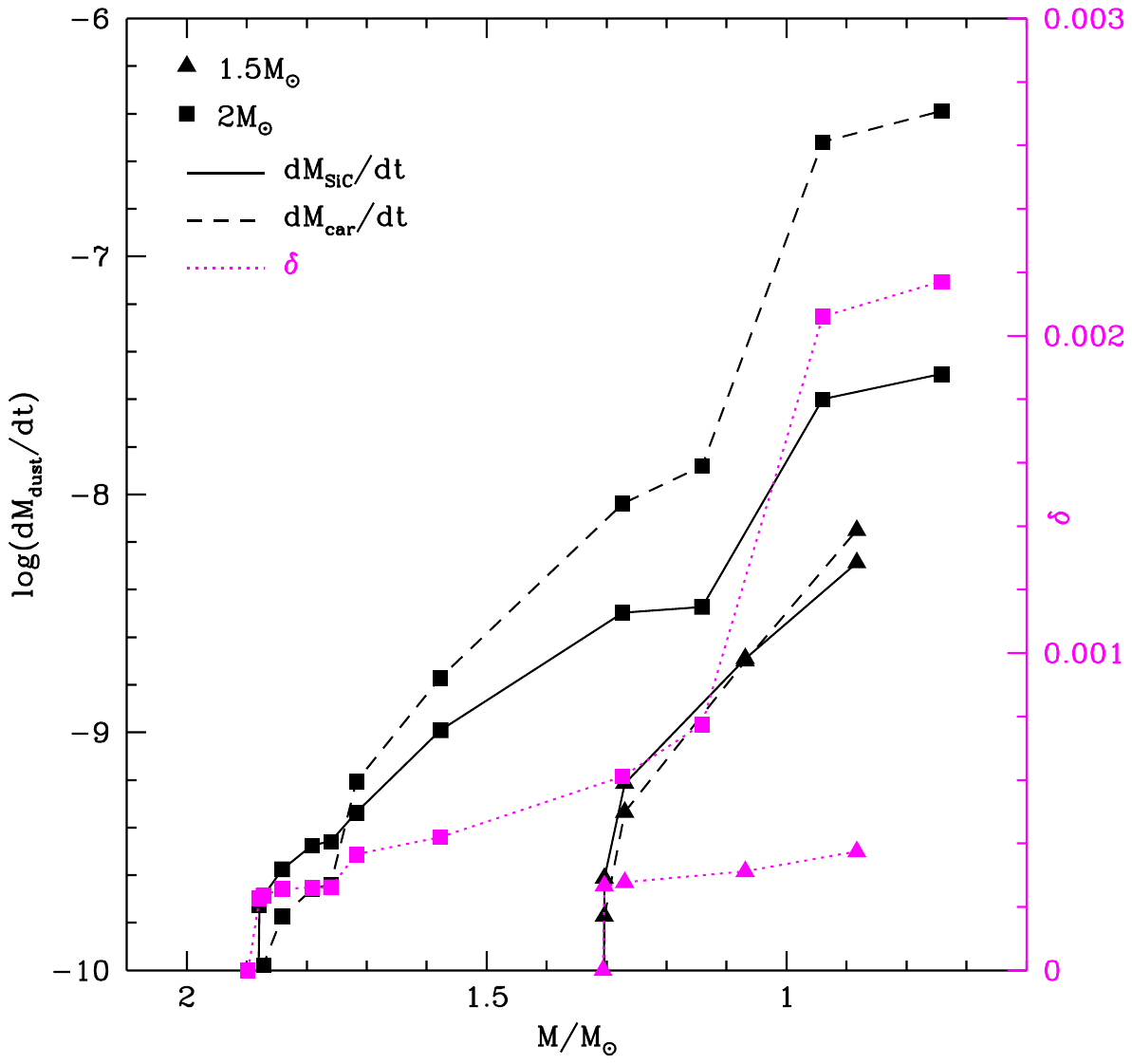}}
\end{minipage}
\vskip-60pt
\caption{Left: time variation of the surface carbon mass fraction (solid lines,
scale on the left) and of the total mass (dashed lines, scale on the right)
of model stars with metallicity $\rm Z=8\times 10^{-3}$, and initial masses
(taken at the start of the core helium burning) $\rm 1.5~M_{\odot}$ (red lines)
and $\rm 2~M_{\odot}$ (black). The first part of the AGB evolution, during
which the stars are oxygen-rich, is indicated with thin lines, whereas the
evolutionary phases after the C-star stage is reached are indicated with thick 
lines. Right: time variation of solid carbon (dashed lines) 
and SiC (solid lines) production rates experienced by the $\rm 1.5~M_{\odot}$ (triangles) 
and $\rm 2~M_{\odot}$ (squares) model stars reported in the left panel, during the
C-star phase. The individual points
refer to the inter-pulse phases. Magenta lines and points 
indicate the variation of the dust-to-gas ratio (scale on the right).
} 
\label{f20}
\end{figure*}

Among the sources studied in paper I, SMP LMC 4 and SMP LMC 71 were identified as
the progeny of $\rm Z=8\times 10^{-3}$ stars of $\rm 1.5~M_{\odot}$ and
$\rm 2~M_{\odot}$, respectively. In Fig.~\ref{f20} we show the evolution of the most
relevant properties of this class of stars, obtained by means of the combination of 
stellar evolution and dust formation modelling, according to the description in
section \ref{input}.

In the left panel of Fig.~\ref{f20} we note the increase in the surface carbon following each TDU event,
the final mass fraction being $\sim 4\times 10^{-3}$ and $\sim 6\times 10^{-3}$
in the $\rm 1.5~M_{\odot}$ and $\rm 2~M_{\odot}$ cases, respectively.
The C-star evolution is limited to the last 3 inter-pulse phases of the
$\rm 2~M_{\odot}$ model star, whereas the $\rm 1.5~M_{\odot}$ model star
evolves as C-star only during the last inter-pulse.
From the evolution of the total mass (dashed lines in the figure) we deduce the notable 
increase in the mass loss
rate that accompanies the achievement of the C-star stage: indeed $\sim 90\%$ of the
envelope mass is lost during the final C-rich phase.

The right panel of Fig.~\ref{f20} regards the dust production properties. For the two model stars considered 
we show the evolution of the dust production rate (split between the solid carbon and the 
SiC components), and the dust-to-gas ratio, estimated as the ratio between the dust formation 
rate and the mass loss rate. Note that in the right panel we use the current mass of the star 
as time indicator, as use of time would shift the most relevant, final part of the evolution,
towards the right side of the figure. 

In the phases immediately following the reaching of the C-star stage the formation of 
SiC is comparable or even more efficient than that of carbon dust, because the carbon
excess with respect to oxygen, which is the key quantity to form carbon dust \citep{fg06}, is
smaller than the gaseous silicon required to form SiC grains. During the AGB evolution
the relative contribution of SiC to the total dust budget diminishes over time due to the ongoing
effects of TDU, which enrich the surface regions in carbon.
For the models considered here we find that the DPR during the last AGB phases is
$\sim 10^{-8}$ and $\rm \sim 5 \times 10^{-7}~M_{\odot}$/yr for the 
$\rm 1.5~M_{\odot}$ and $\rm 2~M_{\odot}$ model stars, respectively; the dust-to-gas
ratios being $\sim 5\times 10^{-4}$ and $\sim 2\times 10^{-3}$.

\begin{figure*}
\vskip-40pt
\begin{minipage}{0.46\textwidth}
\resizebox{1.\hsize}{!}{\includegraphics{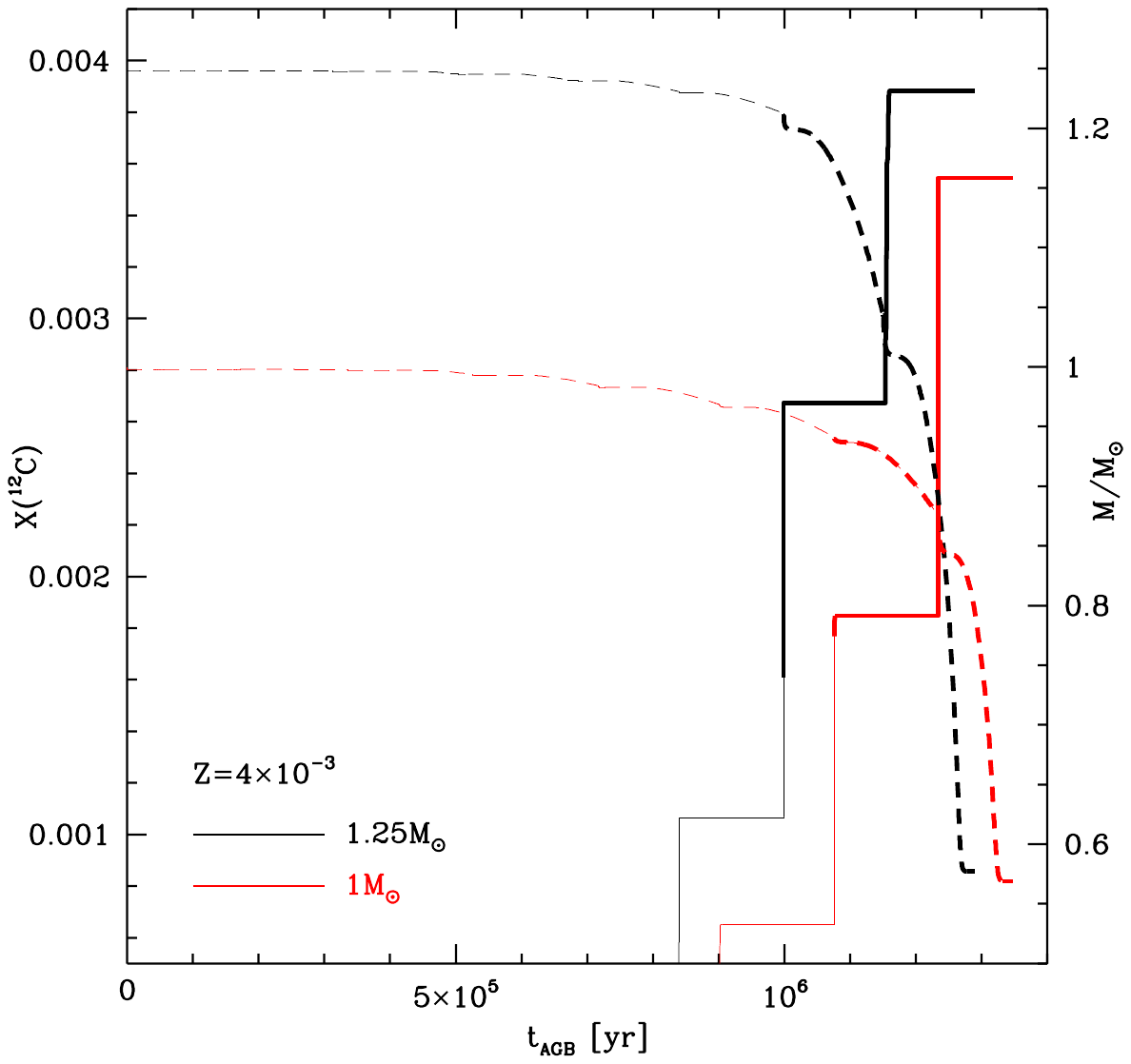}}
\end{minipage}
\begin{minipage}{0.46\textwidth}
\resizebox{1.\hsize}{!}{\includegraphics{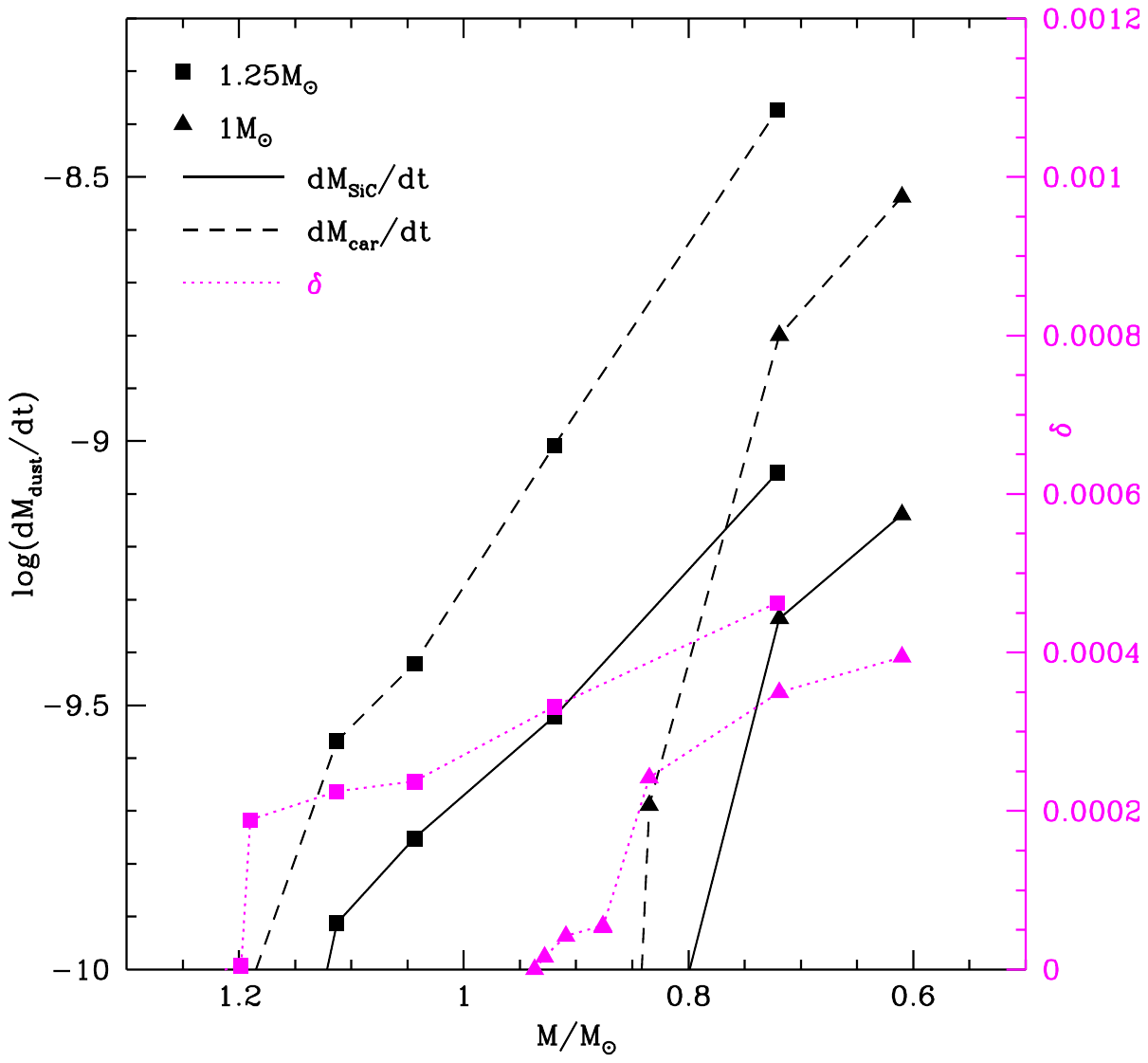}}
\end{minipage}
\vskip-60pt
\caption{Same as Fig.~\ref{f20}, but referring to the evolution of
$\rm 1~M_{\odot}$ (red lines) and $\rm 1.25~M_{\odot}$ (black)
model stars of metallicity $\rm Z=4\times 10^{-3}$. The connection
between the thickness of the lines and the surface $\rm C/O$ is the
same as in Fig.~\ref{f20}
} 
\label{f10}
\end{figure*}

The sources SMP LMC 66 and SMP LMC 102 were identified as the progeny of stars
of metallicity $\rm Z=4\times 10^{-3}$ and mass of $\rm 1~M_{\odot}$ and $\rm 1.25~M_{\odot}$, 
respectively, in paper I. The evolution of these stars is reported in Fig.~\ref{f10},
whose panels show the same quantities as in Fig.~\ref{f20}. In both the model stars
presented in Fig.~\ref{f10} the C-star phase is restricted to the two last inter-pulse
phases. The final surface carbon abundances are in the $3\times 10^{-3}-4\times 10^{-3}$
range, a bit smaller than seen in the previous cases, in agreement with the trend
of final carbon vs initial mass of the star discussed earlier in this section.
The dust production rates are also smaller than in the higher mass model stars
discussed above, the final $\rm \dot M_{dust}$ being $\rm \sim 3\times 10^{-9}~M_{\odot}/$yr and
$\rm \sim 5\times 10^{-9}~M_{\odot}/$yr for the $\rm 1~M_{\odot}$ and $\rm 1.25~M_{\odot}$
model stars, respectively. As far as the dust-to-gas ratio is concerned, we see in the right panel
of Fig.~\ref{f10} and in Table 1 that the final value is $\sim 4\times 10^{-3}$ for the $\rm 1~M_{\odot}$
model star, while it is $\sim 5\times 10^{-3}$ in the $\rm 1.25~M_{\odot}$ case.

\begin{figure*}
\vskip-40pt
\begin{minipage}{0.46\textwidth}
\resizebox{1.\hsize}{!}{\includegraphics{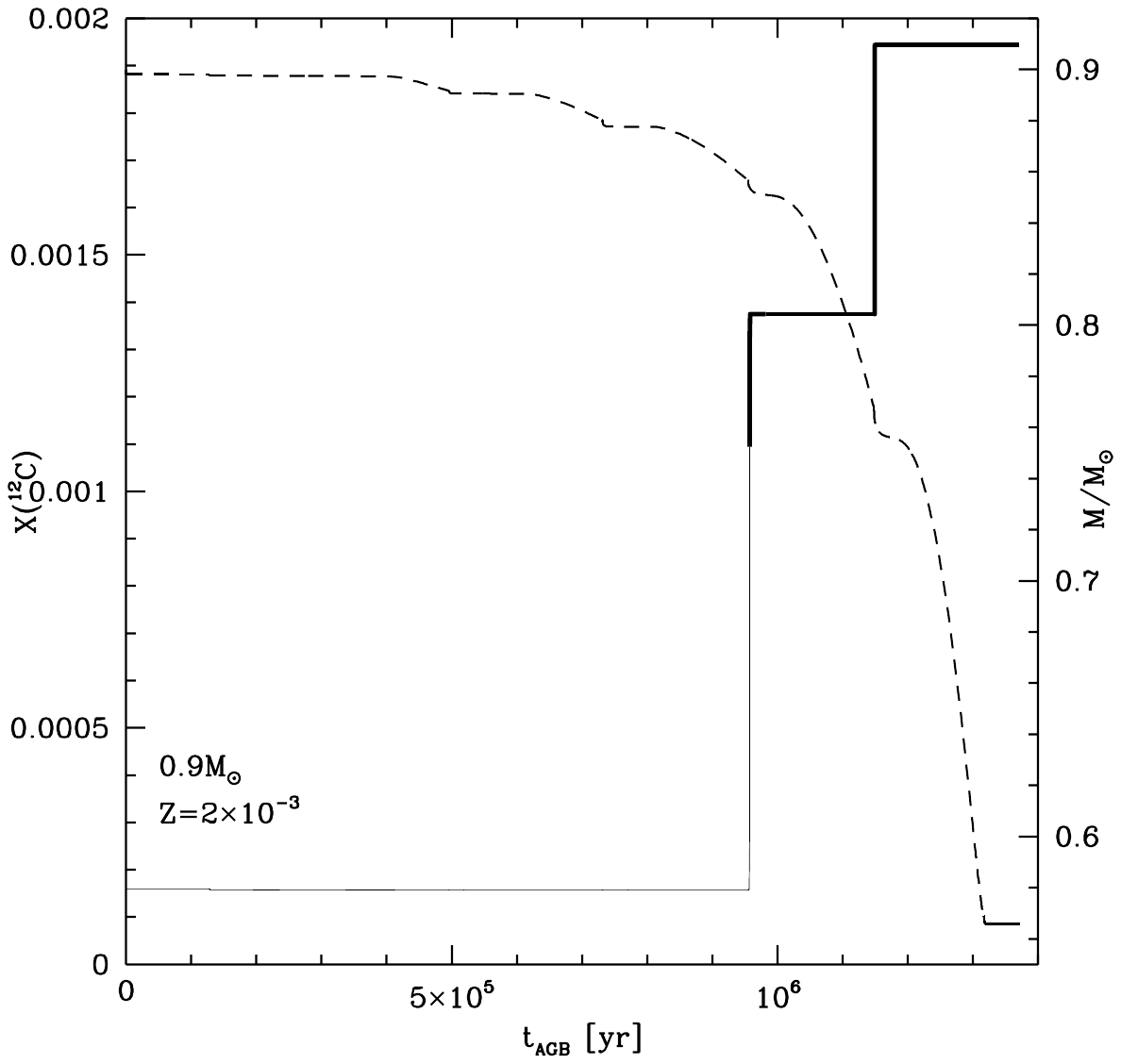}}
\end{minipage}
\begin{minipage}{0.46\textwidth}
\resizebox{1.\hsize}{!}{\includegraphics{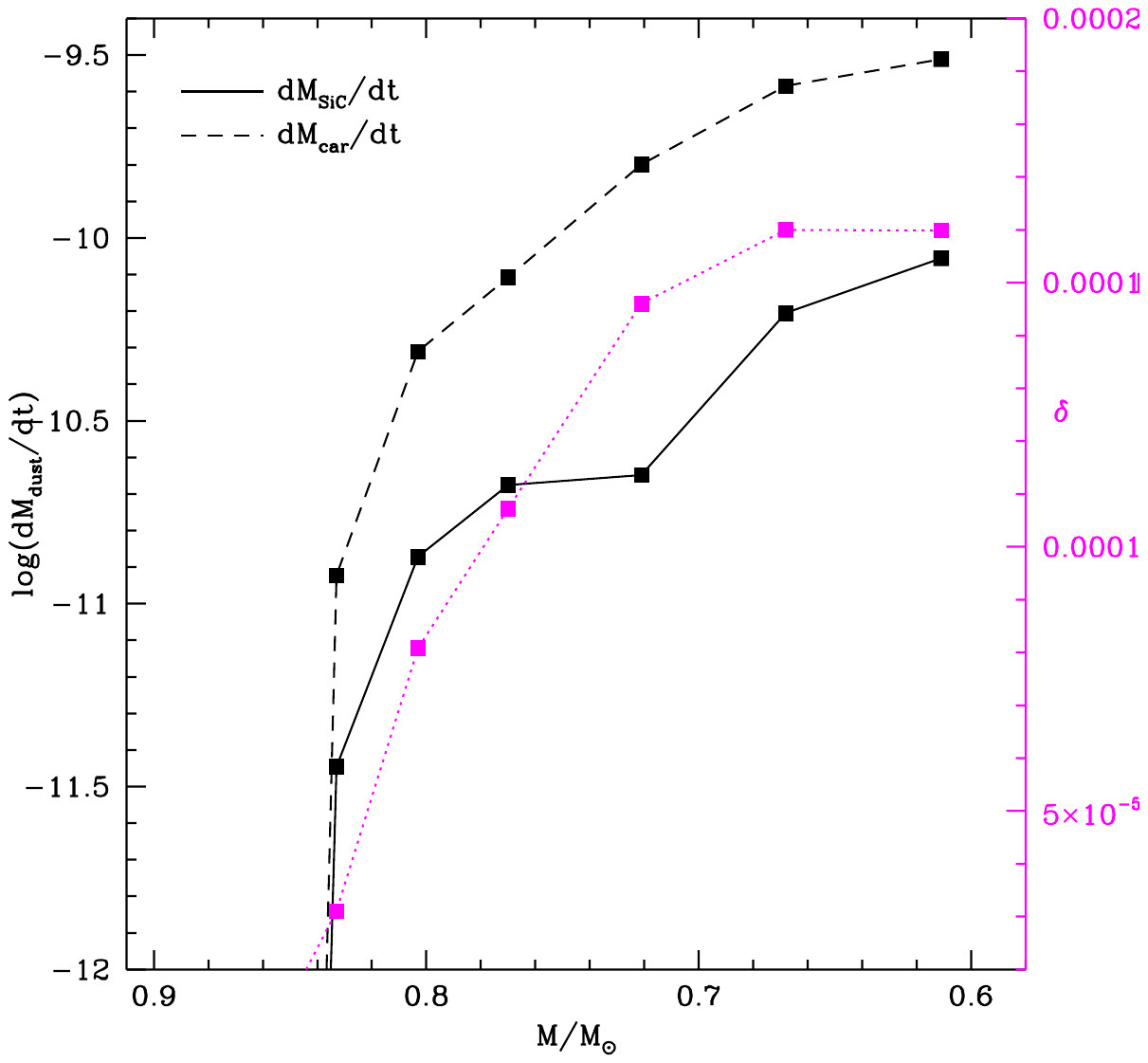}}
\end{minipage}
\vskip-60pt
\caption{Same as Fig.~\ref{f20} and \ref{f10}, but referring to the evolution of
a $\rm 0.9~M_{\odot}$ model star of metallicity $\rm Z=2\times 10^{-3}$
} 
\label{f09}
\end{figure*}

The sources SMP LMC 18, SMP LMC 25 and SMP LMC 34 investigated in paper I
were associated to the evolution of low-metallicity stars ($\rm Z \sim 0.002$) of mass
$\rm \sim 0.9~M_{\odot}$. In the metal-poor domain these masses are just above the minimum
threshold required to reach the C-star stage \citep{devika23}. A typical evolution of such 
a star is shown in Fig.~\ref{f09}, where we report the evolution of the
same quantities as in Fig.~\ref{f20}.
It is evident from the run of the stellar mass that most of the envelope loss takes
place during the last two inter-pulses, during which in fact the star evolves as carbon star.
The final surface carbon is $\sim 2\times 10^{-3}$, whereas the largest rate of dust
production, obtained upon adding the solid carbon and SiC contributions, reached 
at the very end of the AGB evolution, is $\rm \sim 2.5\times 10^{-9}~M_{\odot}/$yr, 
significantly smaller than the values found for higher mass counterparts, as reported in Fig.~\ref{f20} and \ref{f10}.

\begin{figure}
\vskip-40pt
\centering
\begin{minipage}{0.48\textwidth}
\resizebox{1.\hsize}{!}{\includegraphics{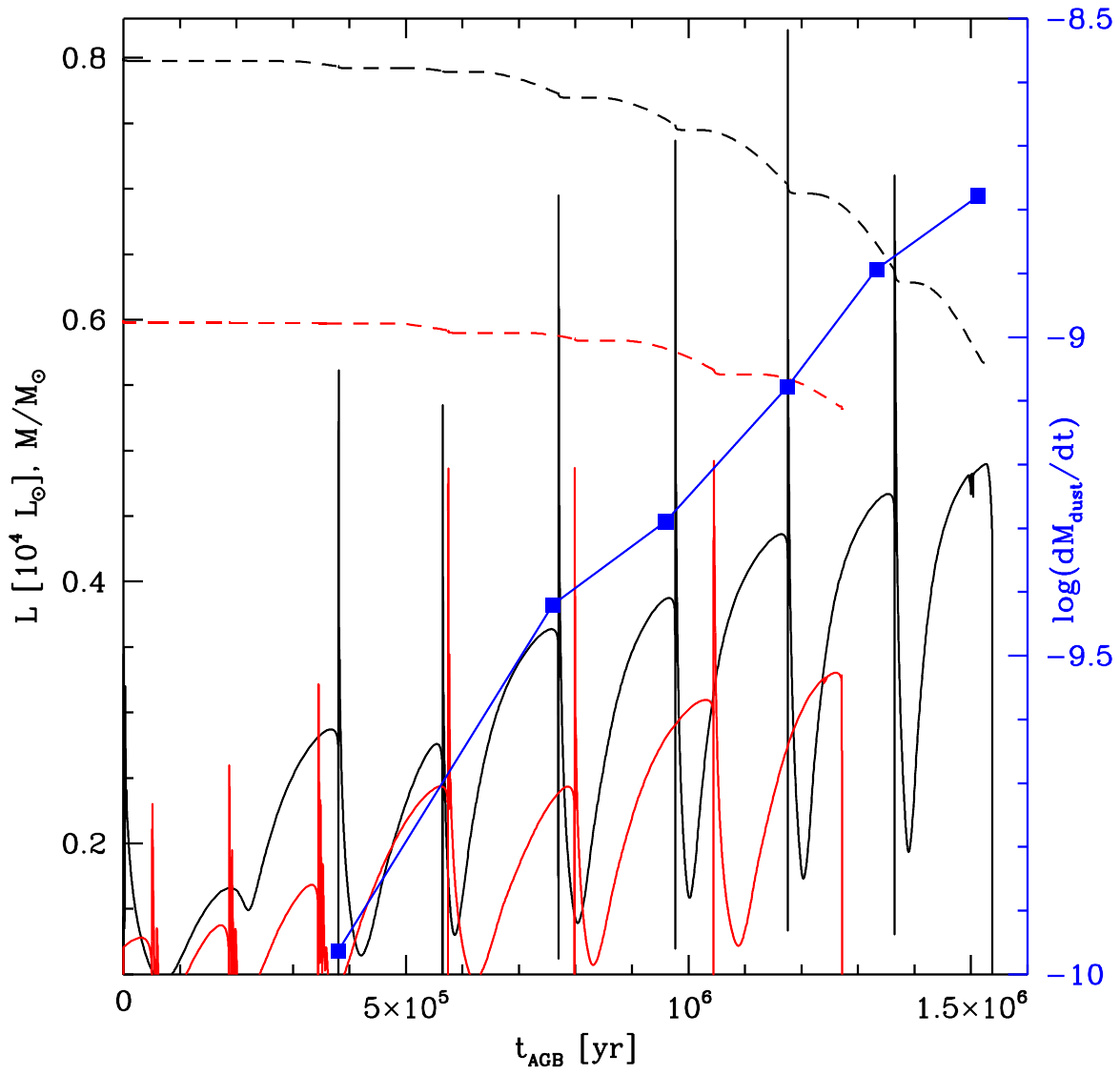}}
\end{minipage}
\vskip-60pt
\caption{Time variation of the luminosity (solid lines) and of the total mass
(dashed) of model stars of metallicity $\rm Z=4\times 10^{-3}$ and initial
masses $\rm 0.6~M_{\odot}$ (red) and $\rm 0.8~M_{\odot}$ (black). The blue
line indicates the DPR of the $\rm 0.8~M_{\odot}$ model star. 
}
\label{f08}
\end{figure}

We conclude this review of the sources studied in paper I with SMP LMC 80 and
SMP LMC 81, which the authors interpreted as the progeny of low-mass stars that
never reached the C-star stage. More specifically, as reported in Table 1,
SMP LMC 81 descends from a $\rm Z=4\times 10^{-3}$ progenitor of mass around 
$\rm 0.8~M_{\odot}$, whereas SMP LMC 80 is the progeny of a $\sim \rm 0.6~M_{\odot}$ 
star of similar metallicity.

Fig.~\ref{f08} reports the main aspects of the evolution of these stars: the time 
evolution of the stellar mass and luminosity are shown for both sources. 
The changes in the dust production rate are shown for the higher mass object only,
as negligible dust production is expected in the case of the lower mass star.

As discussed in \citet{ventura22}, the chemical composition of very low-mass stars 
like SMP LMC 80 and SMP LMC 81 is primarily affected by the first dredge-up process 
(and possible non canonical mixing) occurring during ascending of the RGB, as neither 
HBB nor TDU act during the AGB evolution of the stars in question. The envelope is
lost after only a few thermal pulses (TPs), before they contract and start the post-AGB 
and the PN phase. Dust is produced in limited quantities, the DPR increasing until 
reaching values of the order of a few $\rm 10^{-9}~M_{\odot}/$yr during the final AGB 
phases (see Fig.~\ref{f08}).

\section{Discussion}
\label{disc}
In paper I we adopted the methods described in section \ref{input} to 
determine the physical parameters characterising the nebula surrounding the CS,
listed in Table 1, such as the mass of the gas in the nebula ($\rm M_{gas}$) 
and the dust-to-gas ratio, $\rm \delta_C$\footnote{We use the label $\rm \delta_C$ as 
most of the sources in the sample are C-rich, and one of the two oxygen-rich stars 
shows no evidence of dust. The case of SMP LMC 81, the only oxygen-rich PN with dust, 
will be discussed separately. }

In the previous section we discussed the evolutionary and dust production properties 
of the stars during the AGB phase. We now attempt to relate the current properties
of the PNe considered in paper I with the efficiency of the dust formation process
in the wind of the stars, as they evolve through the very final AGB phases.

\begin{figure}
\vskip-7pt
\centering
\begin{minipage}{0.48\textwidth}
\resizebox{0.95\hsize}{!}{\includegraphics{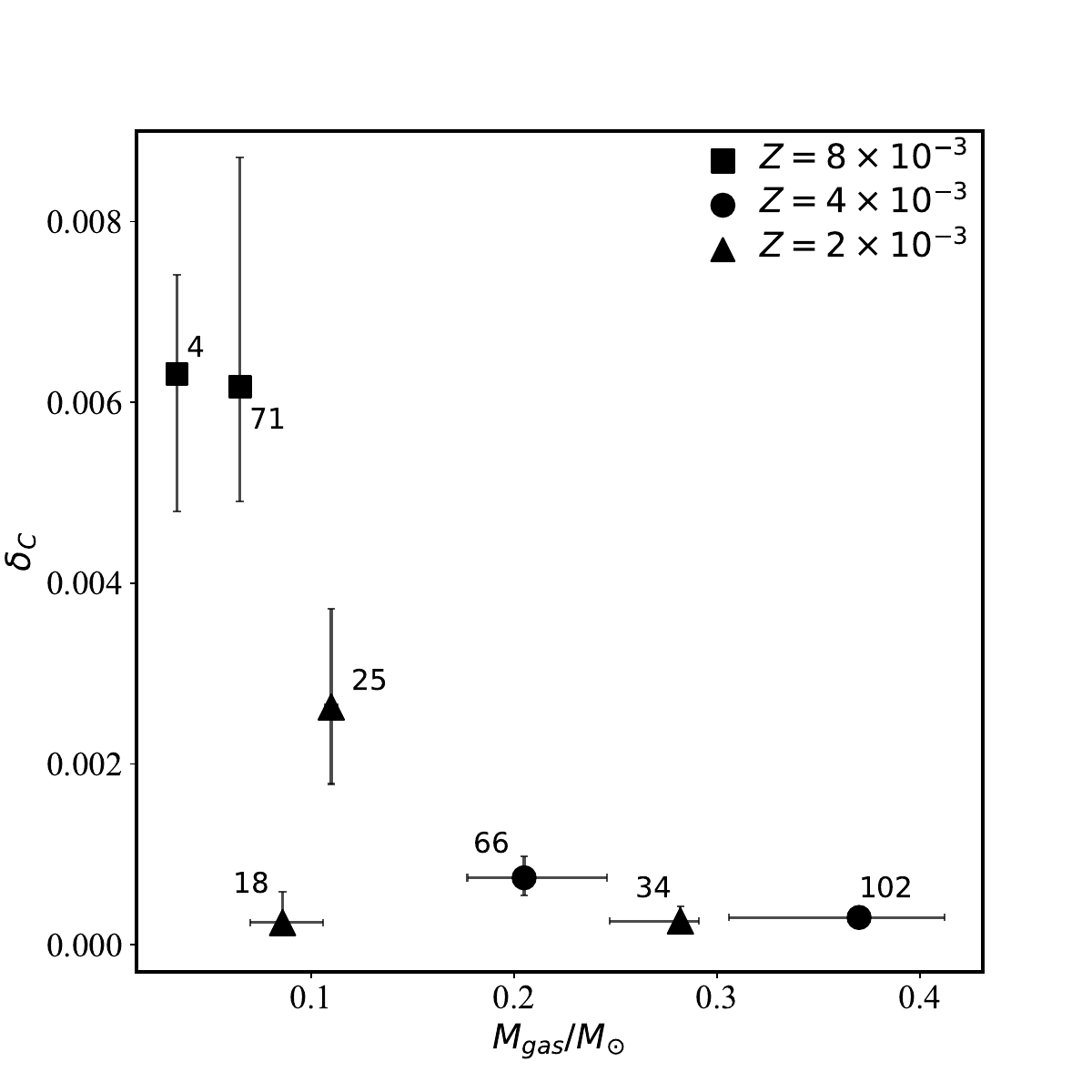}}
\end{minipage}
%\vskip-10pt
\caption{The dust-to-gas ratio derived on the basis of SED fitting for the carbon PNe
investigated in paper I as a function of the mass of the gas in the nebula surrounding the
individual sources. The different symbols refer to the metallicity of the progenitor stars,
determined in paper I on the basis of the position of the sources on the HR diagram and the
surface chemical composition.
}
\label{fall}
\end{figure}

\subsection{The properties of the nebulae}
Fig.~\ref{fall} shows the mass of the gas and the dust-to-gas ratio of the nebulae surrounding 
the carbon stars in the sample investigated here. 
It is clear that the two quantities are anti-correlated: the stars with the largest 
$\rm \delta_C \sim 6\times 10^{-3}$ are surrounded by $\rm \sim 0.05~M_{\odot}$ of gas, 
whereas for the sources with little dust ($\rm \delta_C$ of the order of a few 
$10^{-4}$), $\rm M_{gas}$ is in the $\rm 0.3-0.35~M_{\odot}$ range. 
Overall, the results reported in Fig.~\ref{fall} indicate that the $\rm \delta_C$
values derived from the different sources are related not only to the dust produced during the 
final AGB phases and now stored in the nebula, but also to the fraction of gas lost during the 
AGB-PNe transition.

To discuss how the properties of the PNe are related to the previous history of the stars, 
and how they depend on the mass and chemical composition of the progenitors, we show in 
Fig.~\ref{fdelta} $\rm \delta_C$ (left panel) and the total mass of carbon dust 
nowadays stored in the nebula (right panel) surrounding the different C-rich sources, 
as a function of the mass of the progenitor stars, whose properties were 
discussed in the previous section. The mass of dust was found by multiplying 
$\rm \delta_C$ and $\rm M_{gas}$ deduced on the basis of the analysis of the SED.

The results reported on the left plane indicate that 
$\rm \delta_C$ generally increases with the progenitor's mass, up to the largest
dust fractions, of the order of 0.006.
The masses reported on the abscissa of the right panel of Fig.~\ref{fdelta} 
are the same as in the left panel. For each of the masses considered we also show the
overall dust mass produced during the carbon rich phase, and the dust masses
released during the last one and the last two inter-pulses phases, derived by means
of the dust formation modelling described in section \ref{dustmod}. These quantities
can be deduced based on the time variation of the DPR reported in 
Fig.~\ref{f20}, \ref{f10} and \ref{f09}.

\subsection{Understanding the connection among the dust and gas content of the nebulae, 
and the progenitors' mass}

\citet{tosi22} investigated a sample of post-AGBs belonging to the LMC,
and found that consistency between the observed SEDs and the results from
dust formation and synthetic SED modelling could be obtained under the
assumption that the winds of the stars experiencing very large DPR during the
final AGB phases are faster than those of stars characterised by
a lower efficient dust production at the end of the AGB evolution.
This is consistent with results from dust formation and wind dynamics 
modelling, which demonstrated that the asymptotic velocity of the 
AGB winds increases with the efficiency of the dust formation process,
owing to the enhanced effects of the radiation pressure on dust grains.
This is further confirmed by the modelling of dust formation used to 
obtain the results reported in Fig.~\ref{f20}, \ref{f10} and Fig.~\ref{f09}:
we find that the asymptotic velocity of the outflow during the final AGB phases 
is $\rm \sim 30~km/$s for the $\rm 2~M_{\odot}$ model star, then decreases to 
$\rm \sim 20~km/$s, for the $\rm 1~M_{\odot}$ model star, whereas it is 
slightly below $\rm 10~km/$s in the $\rm 0.9~M_{\odot}$ case.

The $\rm \delta_C$ vs $\rm M_{gas}$ trend shown in Fig.~\ref{fall} can
be explained by considering that the stars populating the left, upper
region of the plane are those that experienced the largest DPR's during the
late AGB evolution, so that they lost a significant fraction of the
gas in their surroundings, owing to the fast winds. Conversely, the
stars in the right, bottom corner descend from progenitors characterised
by poor dust formation during the whole AGB lifetime, so that they were
able to keep most of the gas by the time that they reached the PN stage.

\begin{figure*}
\vskip-15pt
\begin{minipage}{0.46\textwidth}
\resizebox{1.\hsize}{!}{\includegraphics{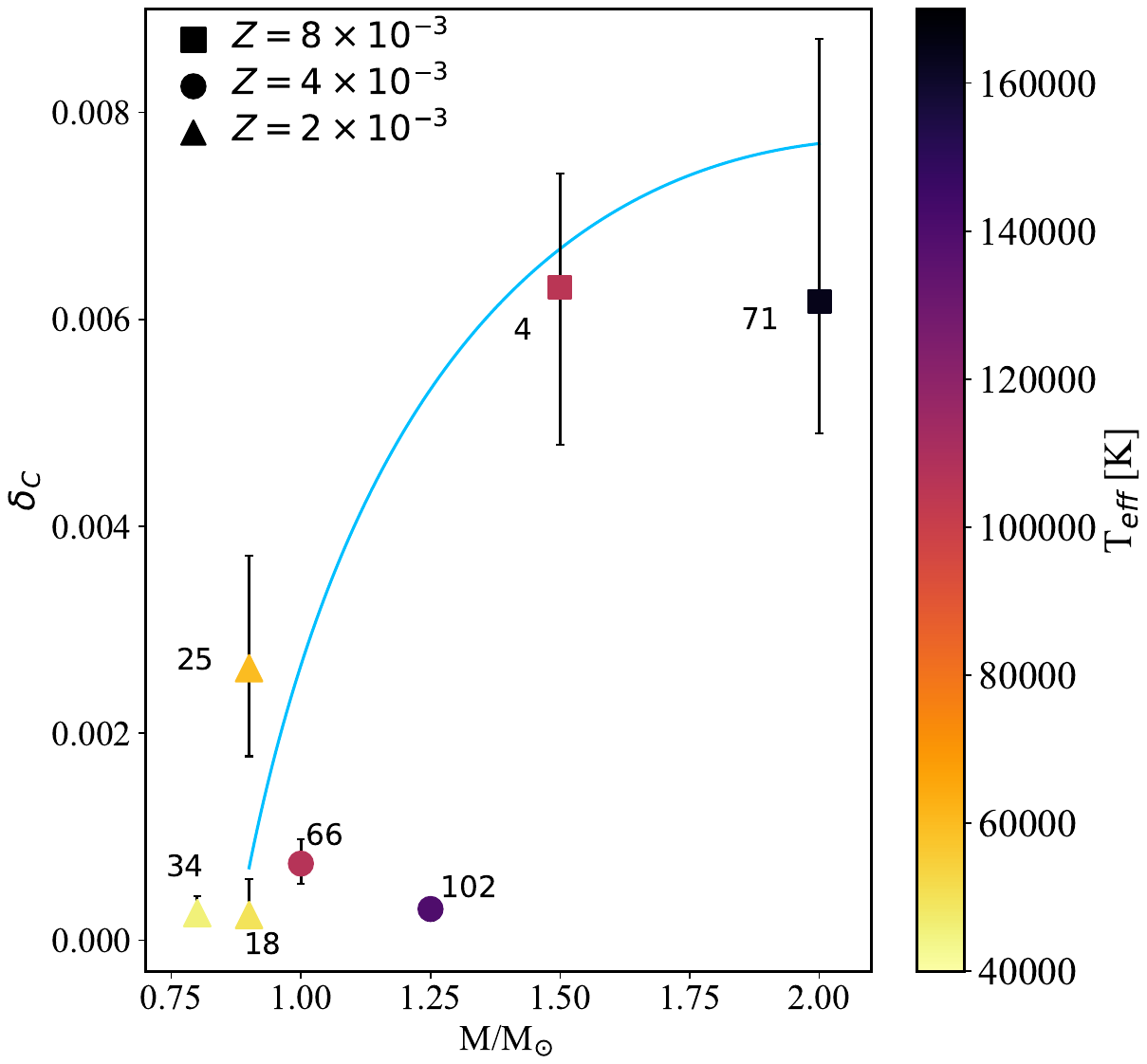}}
\end{minipage}
\begin{minipage}{0.46\textwidth}
\resizebox{1.\hsize}{!}{\includegraphics{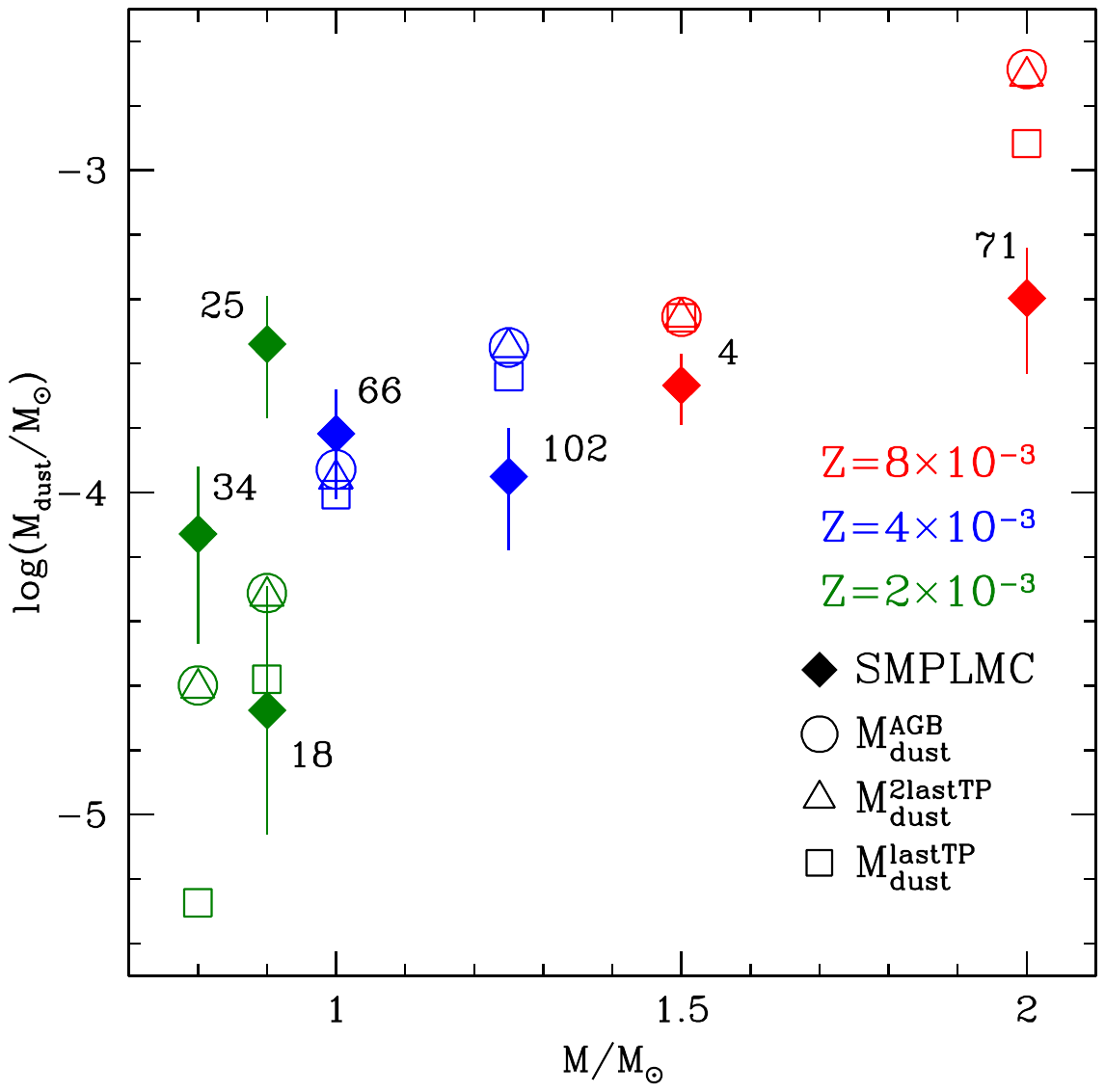}}
\end{minipage}
\vskip-60pt
\caption{Left: The carbon dust-to-gas ratio of the LMC sources
considered in the present work, inferred by SED fitting,
as a function of the progenitors' mass, deduced on the basis of 
the position on the HR diagram and the surface chemical composition.
The cyan line indicates the approximate trend traced by the
sources with effective temperatures below $10^5$ K.
Right: The mass of the dust in the nebula surrounding the individual
sources, as a function of the progenitors's mass, is shown with full 
points. Open points indicate the dust mass produced by model stars
of the same mass during the whole carbon stars phase (circles),
during the late two inter-pulse phases (triangles) and during 
the final inter-pulse phase (squares).
} 
\label{fdelta}
\end{figure*}

The left panel of Fig.~\ref{fdelta} shows that $\rm \delta_C$ generally 
increases with the progenitor's mass, which is a mere consequence of how 
the efficiency of dust production during the very final AGB phases depends 
on the initial mass of the star. Indeed, as discussed in the previous section, 
and shown in the right panels of Fig.~\ref{f20} and \ref{f10}, as well as in 
Fig.~\ref{f09}, the DPR at the tip of the AGB phase increases from 
$\rm \sim 5\times 10^{-9}~M_{\odot}/$yr, for the lowest masses considered here 
($\rm \sim 0.9~M_{\odot}$), to $\rm \sim 5\times 10^{-7}~M_{\odot}/$yr, for 
$\rm M \sim 2~M_{\odot}$. The growing trend of $\rm \delta_C$ with the 
progenitors' mass visible in Fig.~\ref{fdelta} is therefore a consequence of 
the larger dust production in the wind of higher mass carbon stars, 
in turn related to the larger accumulation of carbon during their evolution,
as discussed in section \ref{evol}. A further motivation of the higher 
$\rm \delta_C$ found for the stars of higher mass is that the gas in their surroundings
is dispersed more easily: as discussed earlier in this section, the large DPR's 
enhance the effects of the radiation pressure acting on the newly formed dust 
particles, which leads to faster winds.
It is within this framework that we explain that the sources with the lowest
$\rm M_{gas}$, as clear in Fig.~\ref{fdelta}, are SMP LMC 4 and SMP LMC 71, 
those characterised by the largest DPR during the final AGB phases.
Overall, these results are consistent with the analysis on the PNe sample
of Andromeda presented by \citet{batta}.

\subsection{The relationship between the dust content of the nebulae and the
progenitor's mass}
In the right panel of Fig.~\ref{fdelta} we compare the current estimated mass
of the dust in the nebula with the dust produced by the progenitor stars
during the final AGB phases: in particular, we show the total dust released 
by the stars during the last two inter-pulses experienced, before the start of the
evolution towards the post-AGB phase. These results offer the opportunity
to evaluate the dust survival from the AGB to the current PN phase.

For the stars descending from $\rm M>1~M_{\odot}$ progenitors, the current
dust mass is lower than predicted by dust
formation modelling, an indication that part of the dust was lost, either
driven away by radiation pressure, or vaporised by the hot temperatures in
the nebula. As far as SMP LMC 4 is concerned, the current dust mass is about half 
that found via dust formation modelling, whereas in the cases of SMP LMC 102 and 
SMP LMC 71 the ''dust depletion factor'' is around 3: this is consistent with the
fact that that the latter two stars are the hottest ones in the sample,
with effective temperatures above $10^5$ K, thus we expect that sublimation
of the solid grains destroyed part of the dust in the nebula. We will further 
discuss this point in the final part of the present section.

Moving to lower masses, we find the interesting results regarding SMP LMC 25,
for which the situation is reversed with respect to the larger mass
counterparts: the current dust mass in the nebula, estimated to be 
$\rm \sim 3\times 10^{-4}~M_{\odot}$, is above the overall dust produced during 
the AGB phase, which is found to be below $\rm 5\times 10^{-5}~M_{\odot}$. This 
apparently anomalous behaviour is in fact consistent with the analysis by \citet{tosi22}, 
who outlined that dust production by low-mass, metal-poor AGBs is likely underestimated,
and that higher DPR's were required to reproduce the IR excesses of some metal-poor, 
carbon rich post-AGBs in the Magellanic Clouds. \citet{tosi22} connected this behaviour
with the fact that the recipes used to model dust formation (we refer specifically to
the prescriptions by the Berlin group discussed in section \ref{stevol})
from carbon stars neglect
the role of the surface chemical composition, not taking into account that
lower metallicity stars have a smaller quantity of oxygen, thus the amount of
carbon available to form carbon dust is larger than in the higher metallicity
counterparts, as discussed in section \ref{input}. Indeed the dust mass around SMP LMC 25
is slightly higher than found in the case of SMP LMC 66, thus apparently breaking the 
increasing trend of the dust mass with the progenitor's mass traced by the other stars 
in the sample. According to our reading, this is due to the lower metallicity of
SMP LMC 25, which eases the formation of carbon dust with respect to
SMP LMC 66, despite the higher initial mass of the latter source.

Still in the context of the progeny of low mass stars, we consider the case
of SMP LMC 34, for which the overall carbon dust estimated from the analysis of the
SED is $\rm \sim 2.5\times 10^{-5}~M_{\odot}/$yr, the lowest among the carbon stars
investigated. This is related to the fact that the progenitor's mass is very close
to the minimum threshold mass required to reach the C-star stage, such that the
carbon excess with respect to oxygen, the key quantity for the formation
of solid carbon (see section \ref{input}), is extremely small. Indeed, the results 
from paper I indicate that this source, classified as carbon-rich, can be
considered as a carbon star only if the lower limit of the measured oxygen abundance 
is considered (see Fig.~7 in paper I). The small dust mass and the extremely
low $\rm \delta_C \sim 2.5\times 10^{-4}$ of SMP LMC 34 are a consequence of the
very small carbon excess with respect to oxygen, so that most carbon atoms are locked
into CO molecules, leaving little quantities available to the formation of dust.

SMP LMC 18 is characterised by an extremely small carbon dust mass of 
$\rm \sim 2\times 10^{-5}~M_{\odot}/$yr, when compared to the other carbon stars examined 
here. For this source we discussed earlier in this section the possibility that the dynamics
of the gas and dust in the nebula was altered by the ignition of a late thermal pulse,
which favoured the loss of additional quantities of gas and dust during the transition from 
the post-AGB to the PN phase.

The PNe sample studied in paper I also includes SMP LMC 80 and
SMP LMC 81, which are oxygen-rich. The situation regarding SMP LMC 80 is
straightforward, since the lack of dust in the nebula is explained by the
negligible dust production that characterised the AGB phase of the low-mass
progenitor star, as discussed section \ref{evol}.

On the other hand, the analysis of the SED of SMP LMC 81 done in
paper I showed that some dust is present, with dust-to-gas
ratio $\rm \sim 3.3\times 10^{-3}$, within $\rm \sim 0.13~M_{\odot}$ of gas.
If located into the $\rm M_{gas}$ vs $\rm \delta_C$ plane
in Fig.~\ref{fall}, SMP LMC 81 would fall off the general
sequence traced by the other dusty PNe, as the dust fraction
is higher than the C-rich sources characterised by similar
gas masses. However, comparing these directly may be inappropriate due to intrinsic 
differences between silicate and carbonaceous dust, such as their 
extinction properties and the distance of the dusty layer from the CS.

The analysis of a larger sample 
of oxygen-rich PNe is required before drawing some conclusions
regarding oxygen-rich stars.

\subsection{Some outliers falling off the $\rm M_{gas}-\delta_C$ and
$\rm \delta_C-M/M_{\odot}$ trends}
The only clear deviation from the anti-correlation pattern shown in
Fig.~\ref{fall} is SMP LMC 18, which in paper I 
was interpreted as the descendant of a $\rm \sim 0.9~M_{\odot}$ star that experienced a late
thermal pulse just after the start of the post-AGB phase, thus similar
to the evolution labelled as AFTP by \citet{blocker01}. The ignition of the late
thermal pulse favoured the return of the evolutionary track to the red, followed by
the excursion to the blue, until the present evolutionary stage. This is the reason for 
the relatively low luminosity of SMP LMC 18 $(\rm \sim 2\times 10^3~L_{\odot})$ in 
comparison to the other stars (see the evolutionary track of the $\rm 0.9~M_{\odot}$
of metallicity $Z=2\times 10^{-3}$, in Fig.~5 of paper I). The temporary cooling of 
the external regions following the ignition of the late thermal pulse altered the
dynamics of the gas and dust released during the late AGB phases, likely favouring
the loss of a further amount of residual gas, which is the reason for the
anomalous position of this source in the plane shown in Fig.~\ref{fall}.

The sources SMP LMC 102 and SMP LMC 71 fall out of the main pattern traced 
by the stellar mass vs $\delta_C$ trend, roughly indicated with a cyan line 
in the left panel of Fig.~\ref{fdelta} (actually for SMP LMC 71, when the error
bar is considered, the results are marginally consistent). According to our understanding, this 
is related to the high effective temperatures of their two central stars, 
well above $10^5$ K, which favoured the destruction of an important part of 
the dust present in their surroundings. A similar explanation, although at a 
lower extent, might hold for SMP LMC 66. This interpretation is consistent 
with the previous Spitzer IR observational spectroscopic studies of PNe in 
the Galaxy \citep{letizia12} and in the Magellanic Clouds \citep{letizia07}.
The latter studies show that the large majority of PNe with featureless IR 
spectra (thus no dusty) are among those with the highest effective temperatures 
for their central stars.

\section{Conclusions}
\label{concl}
We use results from stellar evolution and dust formation
modelling to link the properties of the gas and dust in 
a sample of PNe in the LMC, obtained
via SED fitting, with the main parameters of the progenitor 
stars, namely mass and chemical composition. The analysis is 
mostly limited to carbon stars, since the sample considered 
include only two oxygen-rich objects, one of which, descending 
from a very low mass $\rm (\sim 0.6~M_{\odot})$ star, exhibits 
no traces of dust.

We find that, on general grounds, the progenitor's mass
is anti-correlated with the nebular gas mass, while it correlates 
with the dust-to-gas ratio. The dust-to-gas ratio is found to
increase from $\rm \delta_C \sim 5\times 10^{-4}$ to 
$\rm \delta_C \sim 6\times 10^{-3}$, as the progenitor mass increases 
from $\rm 0.9~M_{\odot}$ to $\rm 2~M_{\odot}$. Very hot PNe, with 
effective temperatures hotter than $10^5$ K, fall off this trend,
owing to the strong vaporization at which the dust particles are
exposed in such hot environments.

The understanding of this behaviour is that as far
as C-stars is concerned, higher mass stars produce
more dust during the AGB phase, which per se explains 
the higher dust content in the nebula surrounding the CS.
Furthermore, the larger dust production 
during the late AGB phases of higher mass stars favours 
higher expansion velocities, owing to the enhanced effects 
of radiation pressure on the dust grains: this is
the reason why the gas content of the nebula decreases
from $\rm \sim 0.3-0.35~M_{\odot}$, for $\rm 0.9~M_{\odot}$ progenitors,
to $\rm \sim 0.05~M_{\odot}$, for the descendants
of $\rm 2~M_{\odot}$ stars. This adds a further
justification to the higher dust-to-gas ratios
derived.

This study adds a further contribution to the set up of a methodology 
able to link the properties of PNe with the mass, chemical composition and 
dust production efficiency of the progenitor stars. These information will be used to 
shed new light on the dust production mechanism of low and
intermediate mass stars during the late AGB phases,
which in turn will allow us to understand the role played by AGB stars as dust manufacturers. 
The sample used in the present work must be enlarged to explore the
whole range of the masses of the progenitors of carbon stars
(so far limited to $\rm 2~M_{\odot}$), and must
be extended to oxygen-rich PNe, whose dust is composed by silicates.

\begin{acknowledgements}
This article is based upon work from COST Action CA21126 - Carbon
molecular nanostructures in space (NanoSpace), supported by COST
(European Cooperation in Science and Technology).
PV acknowledges support by the INAF-Theory-GRANT 2022 "Understanding mass loss and dust production from evolved stars".
DAGH acknowledges the support from the State Research Agency (AEI) of
the Ministry of Science, Innovation and Universities (MICIU) of the
Government of Spain, and the European Regional Development Fund (ERDF),
under grants PID2020-115758GB-I00/AEI/10.13039/501100011033 and
PID2023-147325NB-I00/AEI/10.13039/501100011033.
FDA acknowledges suppor by the INAF-Mini-GRANTS 2023 "Understanding evolved stars and their dust production through the lens of planetary nebulae"
DK acknowledges funding support from the Australian Research Council Discovery Project DP240101150.
M.A.G.-M. acknowledge to be funded by the European Union (ERC, CET-3PO, 101042610). Views and opinions expressed are however those of the author(s) only and do not necessarily reflect those of the European Union or the European Research Council Executive Agency. Neither the European Union nor the granting authority can be held responsible for them.
\end{acknowledgements}


\end{document}